\documentclass[aps,prb,reprint,preprintnumbers,superscriptaddress,amsmath,amssymb,bibnotes,longbibliography]{revtex4-1}
\usepackage{graphicx}
\usepackage{dcolumn}
\usepackage{bm}

\usepackage{amssymb}
\usepackage{amsmath}
\usepackage{epstopdf}
\usepackage{booktabs}
\usepackage{threeparttable}
\usepackage[pdfborder=001,colorlinks,linkcolor=blue,anchorcolor=blue,citecolor=blue,urlcolor=blue,filecolor=blue,menucolor=blue,runcolor=blue]{hyperref}
\usepackage{lineno}

\begin{document}

\title{Magnetotransport and electronic structure of the antiferromagnetic semimetal YbAs}

\author{W. Xie}
\affiliation{Center for Correlated Matter and Department of Physics, Zhejiang University, Hangzhou 310058, China}
\author{Y. Wu}
\affiliation{Center for Correlated Matter and Department of Physics, Zhejiang University, Hangzhou 310058, China}
\author{F. Du}
\affiliation{Center for Correlated Matter and Department of Physics, Zhejiang University, Hangzhou 310058, China}
\author{A. Wang}
\affiliation{Center for Correlated Matter and Department of Physics, Zhejiang University, Hangzhou 310058, China}
\author{H. Su}
\affiliation{Center for Correlated Matter and Department of Physics, Zhejiang University, Hangzhou 310058, China}
\author{Y. Chen}
\affiliation{Center for Correlated Matter and Department of Physics, Zhejiang University, Hangzhou 310058, China}
\author{Z. Y. Nie}
\affiliation{Center for Correlated Matter and Department of Physics, Zhejiang University, Hangzhou 310058, China}
\author{S.-K. Mo}
\affiliation{Advanced Light Source, E. O. Lawrence Berkeley National Lab, Berkeley, California 94720, United States}
\author{M. Smidman}
\email{msmidman@zju.edu.cn}
\affiliation{Center for Correlated Matter and Department of Physics, Zhejiang University, Hangzhou 310058, China}
\author{C. Cao}
\affiliation{Department of Physics, Hangzhou Normal University, Hangzhou 310036, China}
\author{Y. Liu}
\email{yangliuphys@zju.edu.cn}
\affiliation{Center for Correlated Matter and Department of Physics, Zhejiang University, Hangzhou 310058, China}
\affiliation{Collaborative Innovation Center of Advanced Microstructures, Nanjing University, Nanjing 210093, China}
\author{T. Takabatake}
\affiliation{Center for Correlated Matter and Department of Physics, Zhejiang University, Hangzhou 310058, China}
\affiliation{Department of Quantum Matter, AdSM, Hiroshima University, Higashi-Hiroshima 739-8530, Japan}
\author{H. Q. Yuan}
\email{hqyuan@zju.edu.cn}
\affiliation{Center for Correlated Matter and Department of Physics, Zhejiang University, Hangzhou 310058, China}
\affiliation{Collaborative Innovation Center of Advanced Microstructures, Nanjing University, Nanjing 210093, China}

\date{\today}

\begin{abstract}
A number of rare-earth monopnictides have topologically non-trivial band structures together with magnetism and strong electronic correlations. In order to examine whether the antiferromagnetic (AFM) semimetal YbAs ($T\rm_N$ = 0.5 K) exhibits such a scenario, we have grown high-quality single crystals using a flux method, and characterized the magnetic properties and electronic structure using specific heat, magnetotransport and angle-resolved photoemission spectroscopy (ARPES) measurements, together with density functional theory (DFT) calculations. Both ARPES and DFT calculations find no evidence for band inversions in YbAs, indicating a topologically trivial electronic structure. From low-temperature magnetotransport measurements, we map the field-temperature phase diagram, where we find the presence of a field stabilized phase distinct from the AFM phase at low temperatures. An extremely large magnetoresistance (XMR) for both YbAs and the nonmagnetic counterpart LuAs, is also observed, which can consistently be accounted for by the presence of electron-hole compensation. Moreover, an angle-dependent study of the Shubnikov-de Haas effect oscillations reveals very similar Fermi surfaces between YbAs and LuAs, with light effective masses down to at least 0.5 K, indicating that the Yb-4\emph{f} electrons are well localized, and do not contribute to the Fermi surface. However, the influence of the localized Yb-4\emph{f} electrons on the magnetotransport of YbAs can be discerned from the distinct temperature dependence of the XMR compared to that of LuAs, which we attribute to the influence of short-ranged spin correlations that appear well above $T\rm_N$.

\begin{description}
\item[PACS number(s)]

\end{description}
\end{abstract}

\maketitle

\section{\label{sec:Introduction}Introduction}

Searching for correlated and magnetic materials with topologically non-trivial electronic structures has recently led to the discovery of novel topological phases, such as topological Kondo insulators \cite{2010TKI,2013SmB6}, magnetic topological insulators \cite{2010AFM-TI,2019MnBi2Te4,2019MnBi2Te4-1,2019MnBi2Te4-2}, magnetic Weyl semimetals \cite{2016TRSB-WF,Co3Sn2S2-1,Co3Sn2S2-2,Co3Sn2S2-3} and Weyl-Kondo semimetals \cite{2017Qimiao,Guo-YbPtBi}. The $LnX$ series crystallizing in the cubic NaCl-type structure \cite{structure}($Ln$ = rare-earth and \emph{X} = As, Sb or Bi) is a large family of semimetals where topologically nontrivial band structures have been found to exist with electronic correlations and magnetism \cite{11-2017LaAs,11-2015LaSb,11-2016LaSb,11-2016NJP-LaBi,11-2016LaBi,11-2017LaBi,11-2017LuSb,11-2018HoSb,11-2018HoBi,11-2018npj, 11-2017CeSb,11-2016NdSb,11-2016JPCM-NdSb,11-2018NdSb,11-2017PrSb, 11-2018SmSb, 11-2018PLi,11-2019Wu, Cao}. For example, evidence was found that CeSb  hosts Weyl fermions in the field-induced ferromagnetic state \cite{11-2017CeSb}, while NdSb was reported to be a Dirac semimetal with antiferromagnetic (AFM) order \cite{11-2016NdSb,11-2016JPCM-NdSb,11-2018NdSb}. Meanwhile in SmSb, the analysis of the Shubnikov-de Haas (SdH) oscillations indicates a topologically non-trivial band structure \cite{11-2018SmSb}. Topological electronic structures were detected using angle-resolved photoemission spectroscopy (ARPES) for several rare-earth mono-bismuthides (Ce, Pr, Sm, Gd)Bi, showing tunable bulk band inversions and corresponding surface states \cite{11-2018PLi}, where the ARPES results agree well with density functional theory (DFT) calculations assuming the $4f$-electrons to be well localized  \cite{Cao}.

Yb-monopnictides are also promising candidates for studying topological systems with electronic correlations and magnetism, but the topology of the electronic structures has not been reported. In this work we focus on one example YbAs, which is an antiferromagnet with a N\'{e}el temperature \emph{T}$\rm_N$ = 0.5 K \cite{ott1985}. Heavy fermion behavior was reported from several results, including a greatly reduced entropy of 0.2\emph{R}$\ln$2 released at $T\rm_N$ together with a reduced ordered moment of 0.85$\mu$$_B$/Yb \cite{ott1985,Donni1989,Bonville1988,1988Bonville et al},  and a huge value of the nuclear relaxation rate $1/(T_1T)$ below 0.2~K \cite{1998NMR}. On the other hand, the resistivity $\rho(T)$  shows metallic behavior, without a logarithmic temperature dependence or Kondo coherence peak \cite{1990transport}, and  the carrier concentration $n\sim$ 10$^{20}$cm$^{-3}$ is only about 1\%  of the magnetic Yb-ions, which is far less than that required for full Kondo screening \cite{1990transport}. A de Haas-van Alphen (dHvA) effect study revealed the presence of three ellipsoidal electron pockets and one spherical hole pocket, where the charge carriers all have rather small effective masses \cite{1992dHvA}.

Various scenarios have been proposed to account for the seemingly contradictory experimental results \cite{1994Kasuya,1995Kasuya,1997Kasuya,1988Takagi,1990transport,1988Bonville, 1988Bonville et al}, among which the \emph{p}-\emph{f} mixing model has been widely discussed \cite{1988Takagi,1990transport,1992dHvA,Takuo1992}. In this scenario, the hole states become heavy, but the electrons remain light and mobile, and therefore the metallic behavior of  $\rho(T)$ is explained by electrons dominating the transport properties, while the additional hole pockets not detected in previous dHvA studies are expected to have large effective charge carrier masses, accounting for the greatly enhanced Sommerfeld coefficient \cite{1992dHvA,Takuo1992}. On the other hand, inelastic neutron scattering (INS) measurements reveal the development of short-range spin correlations below around 20 K  \cite{1995short range,1998short range}, which may also explain the reduced magnetic entropy at $T\rm_N$, which is gradually released up to 20 K where the spin correlations disappear.  As a result,  the reduced ordered moment was proposed to arise due to the coexistence of short-range correlations with long-range magnetic order \cite{1995short range}.

In this work we address several issues by studying high-quality single crystals of YbAs. We firstly characterize the low-temperature magnetism  by determining the response of the magnetic transition to applied magnetic fields, and we construct a temperature--field phase diagram. Moreover, to gain insight into the effect of correlations as well as the topology of the band structure, the Fermi surfaces and carrier effective masses were investigated via angle dependent SdH oscillations, ARPES measurements and DFT calculations. SdH oscillations reveal five Fermi surfaces pockets, one more than previous studies \cite{1992dHvA}, and an analysis of the temperature dependence of the SdH amplitudes above $T_{\rm N}$ shows that the charge carriers of all pockets have light effective masses. Meanwhile, no evidence for a Kondo resonance was found from ARPES measurements down to 10~K, and the results are in good agreement with DFT calculations assuming well localized 4\emph{f}-electrons. Furthermore, these demonstrate the absence of band inversions in the electronic structure, and therefore a trivial band topology. On the other hand, an extremely large magnetoresistance is observed at low temperatures, which from the Kohler's rule scaling analysis and Hall measurements, can be ascribed to electron-hole compensation and large mobilities. Additional features are observed in the low temperature XMR of YbAs not present in the data of LuAs, which are likely due to the influence of short-ranged spin correlations on the magnetotransport.

\section{\label{sec:exp} Experimental details}

\begin{figure}[!h]
\begin{center}
  \includegraphics[width=0.9\columnwidth]{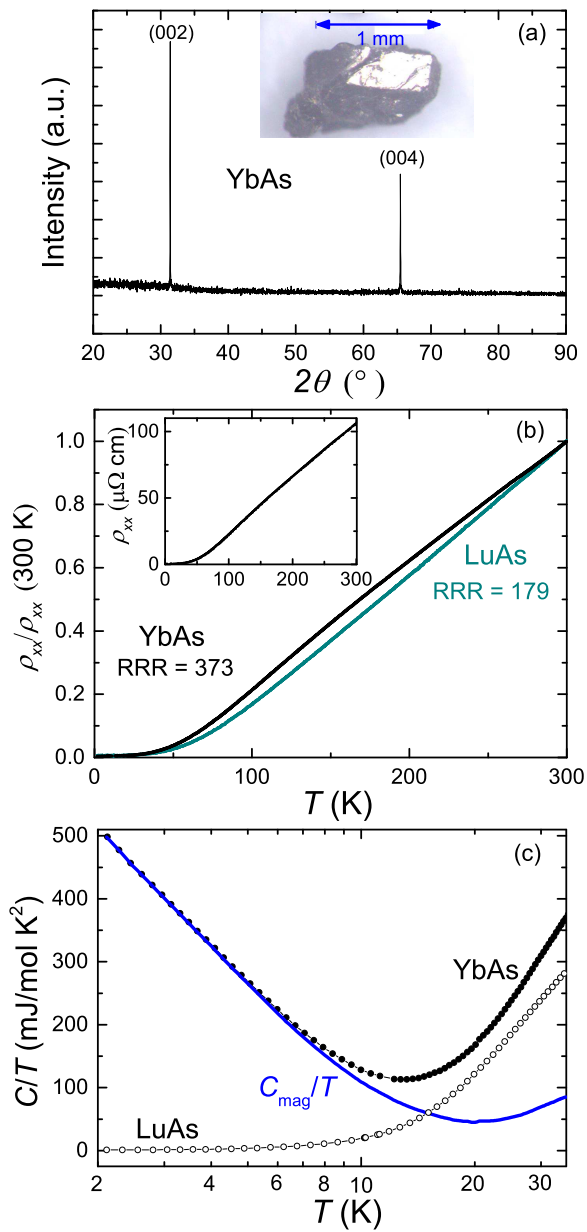}
\end{center}
	\caption{(Color online) (a) XRD pattern on an YbAs single crystal surface with the (001) plane. The inset shows a piece of single crystal with typical dimensions labelled. (b) Normalized temperature dependent resistivity $\rho_{xx}$(\emph{T}) for YbAs and LuAs in zero applied field. The inset of (b) displays the absolute value of $\rho_{xx}$(\emph{T}) of YbAs, where there is a small residual resistivity of 0.27 $\mu\Omega$ cm at 1.9~K. (c) Low-temperature heat capacity as \emph{C}/\emph{T} of YbAs and LuAs. The blue solid line displays the magnetic contribution to the heat capacity of YbAs.}
   \label{figure1}
\end{figure}

Single crystals of YbAs and LuAs were prepared using eutectic Pd-As binary phases as a flux, in contrast to previous studies where samples were grown using the Bridgman method \cite{1999Okayama}. Yb (or Lu) powder, Pd powder and prereacted PdAs$_2$ were placed in an alumina crucible in the molar ratio 1:2:2. The crucible was sealed in an evacuated quartz tube. The mixture was slowly heated to 1000$^\circ$C in a muffle furnace and held there for one day. The furnace was then slowly cooled to 800$^\circ$C at a rate of about 1.8$^\circ$C/h, before the quartz tube was centrifuged to separate the crystals from the flux. Single crystals with a typical size of about 1$\times$1$\times$1 mm$^3$ were obtained, as shown in the inset of Fig.~\ref{figure1}(a).

The crystal structure and orientation were confirmed using x-ray diffraction (XRD) on a cleaved crystal using a PANalytical X'Pert MRD powder diffractometer with Cu \emph{K$_{\alpha1}$} radiation monochromated by graphite. The chemical composition was determined by energy-dispersive x-ray spectroscopy using a Hitachi SU-8010 Field Emission scanning electron microscope. The resistivity and angle-dependent magnetotransport were measured using a Quantum Design Physical Property Measurement System (PPMS) with a sample rotation option. Four Pt wires were attached to the sample using silver paste. The specific heat was also measured using a PPMS. The low-temperature MR between 0.275~K and 10~K was measured using a $^3$He system in fields up to 15 T. The Hall measurements were performed using a PPMS, utilizing a four-wire-method on a piece of single crystal with the applied field perpendicular to the (111) plane and the current direction within the plane, where the transverse resistivity contribution was subtracted.

High-resolution ARPES measurements were carried out on the beamline 10.0.1 at the Advanced Light Source (ALS), Lawrence Berkeley National Laboratory (USA). The typical energy and angular resolution were about 15 meV and 0.2$^\circ$, respectively. A photon energy range from 30 eV to 200 eV was used during the measurements. Single crystals of YbAs were cleaved \emph{in situ} at around 20 K  and measured at the same temperature. Temperature-dependent scans were performed from 13 K up to 80 K, but no obvious temperature dependence was found in the ARPES spectra.  All measurements were performed in an ultrahigh vacuum with a base pressure lower than 1$\times$10$^{-10}$ torr.

Density functional theory (DFT) calculations were performed using the Vienna ab-initio simulation package (VASP). A plane-wave basis up to 380 eV and 12$\times$12$\times$12 $\Gamma$-centered K-mesh were used to make sure the total energy converges to 1 meV per unit cell. The modified Becke-Johnson (mBJ) potentials were used with the effect of spin-orbit coupling (SOC) considered. The \emph{f}-electrons are treated as core states in all the calculations, which gives good agreement with experimental results.

\section{\label{sec:results} Results and discussion}

\subsection{Sample characterization}

 Figure~\ref{figure1}(a) shows an XRD pattern of a cleaved single crystal, in which the diffraction peak positions correspond to the (00$l$) plane of YbAs, indicating the correct phase of our crystals with the [001] direction perpendicular to the cleaved face. Figure~\ref{figure1}(b) shows the normalized temperature dependence of the resistivity $\rho_{xx}(T)$/$\rho_{xx}$(300 K) from 300~K to 1.9~K for YbAs and LuAs. Both display metallic behavior, and at elevated temperatures there is only a slight enhancement of the YbAs data over that of LuAs.
 The absolute resistivity of YbAs is shown in the inset, where the residual resistivity at 1.9~K is about 0.27~$\mu\Omega$ cm. For YbAs, the residual resistivity ratio RRR (= $\rho_{xx}$(300 K)/$\rho_{xx}$(1.9 K)) of 373 is more than one order of magnitude higher than previous studies \cite{1999Okayama}. The RRR of LuAs is about 179, which is also higher than that reported for polycrystalline samples \cite{2014LuAs}. The specific heat as $C/T$ of LuAs monotonically decreases with temperature, and from fitting the low temperature data using $C/T = \gamma_\mathrm{n} + \beta T^2$, a Sommerfeld coefficient of $\gamma\rm_n$ = 0.15 mJ/mol K$^2$ and $\beta$ = 0.14 mJ/(mol-K)$^4$ are obtained, where the latter corresponds to a Debye temperature of 302 K. In contrast, the $C/T$ data of YbAs reaches a minimum at about 15 K, below which the data continues to increase with decreasing temperature.
  The magnetic contribution to the specific heat $C_\mathrm{mag}/T$ (Fig.~\ref{figure1}(c)), obtained by subtracting the $C/T$ of LuAs from that of YbAs, upturns below 20 K with a more rapid increase below about 8 K. This points to a significant low temperature contribution arising due to the presence of the Yb-4\emph{f} electrons.

\begin{figure}[t]
\begin{center}
  \includegraphics[width=\columnwidth,keepaspectratio]{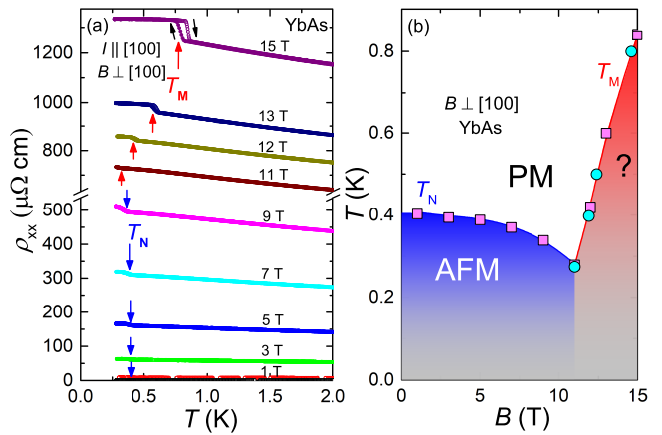}
\end{center}
	\caption{(Color online) (a) Temperature dependence of the resistivity ($\rho$$_{xx}$) of YbAs below 2~K measured in applied magnetic fields up to 15~T. The blue and red vertical arrows denote the transitions at \emph{T}$\rm_N$ and \emph{T}$\rm_M$, respectively. (b) Temperature-field phase diagram of YbAs constructed from $\rho_{xx}$(\emph{T}) and $\rho_{xx}$(\emph{B}). The labels PM and AFM correspond to the paramagnetic phase, and antiferromagnetic phase, respectively, while the nature of the phase below \emph{T}$\rm_M$ is not yet determined.  }
   \label{figure2}
\end{figure}

\subsection{\emph{B}--\emph{T} phase diagram}

\begin{figure}[t]
\begin{center}
  \includegraphics[width=0.9\columnwidth]{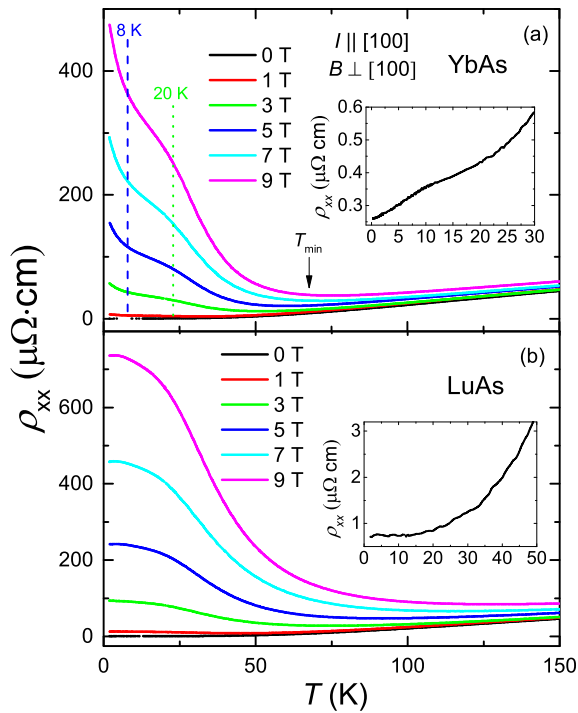}
\end{center}
	\caption{(Color online) Temperature dependence of the resistivity $\rho$$_{xx}$ under various applied magnetic fields down to 1.9~K for (a) YbAs and (b) LuAs, with $\emph{B}$ $\bot$ $\emph{I}$ $\|$ [100]. Two characteristic temperatures are marked for YbAs. The insets of (a) and (b) show the low-temperature part of the zero field resistivity of YbAs and LuAs down to 0.275 K, respectively.}
   \label{figure3}
\end{figure}

From low-temperature resistivity measurements in applied transverse fields, we determined the field dependence of \emph{T}$\rm_N$. Although $\rho$$_{xx}(T, 0)$ shows no detectable anomaly at \emph{T}$\rm_N$ = 0.5 K (inset of Fig.~\ref{figure3}(a)), clear jumps can be resolved in the in-field data. As shown in Fig.~\ref{figure2}(a), $\emph{T}$$\rm_N$ decreases with applied field up to 9~T, while above 11~T, a more pronounced jump appears in $\rho$$_{xx}$ (denoted as $\emph{T}$$\rm_M$), which moves to higher temperature with increasing field. The solid black arrows in Fig.~\ref{figure2}(a) denote the cooling and warming processes of the resistivity measurements. While an negligibly small thermal loop is found at $\emph{T}$$\rm_N$, above 11~T a sizable loop occurs at $\emph{T}$$\rm_M$, which becomes larger with increasing field. Note that the magnitude of the anomaly at the transition increases with field for both $\emph{T}$$\rm_N$ and $\emph{T}$$\rm_M$. Based on $\rho$$_{xx}$(\emph{T}) and $\rho$$_{xx}$(\emph{B}), the $\emph{B}$--$\emph{T}$ phase diagram of YbAs was constructed, as shown in Fig.~\ref{figure2}(b).

A plausible explanation for the first order transition at $\emph{T}$$\rm_M$, which becomes increasingly stable with increasing magnetic field is that this corresponds to the onset of quadrupolar ordering. This is supported by previous reports of a strong quadrupolar interaction in YbAs \cite{1994Kreller, magnetocaloric}, where the quadrupolar moments could originate from the mixture of the CEF ground state $\Gamma_6$ and the first excited CEF state $\Gamma_8$ under applied magnetic fields. Other examples of field-induced quadrupolar ordering in rare earth compounds include Dy$_3$Ru$_4$Al$_{12}$ \cite{Dy3Ru4Al12} and HoFe$_2$Al$_{10}$ \cite{HoFe2Al10}. However, this proposal is to be confirmed by further studies, such as ultrasound measurements.

\begin{figure}[t]
\begin{center}
  \includegraphics[width=0.9\columnwidth,keepaspectratio]{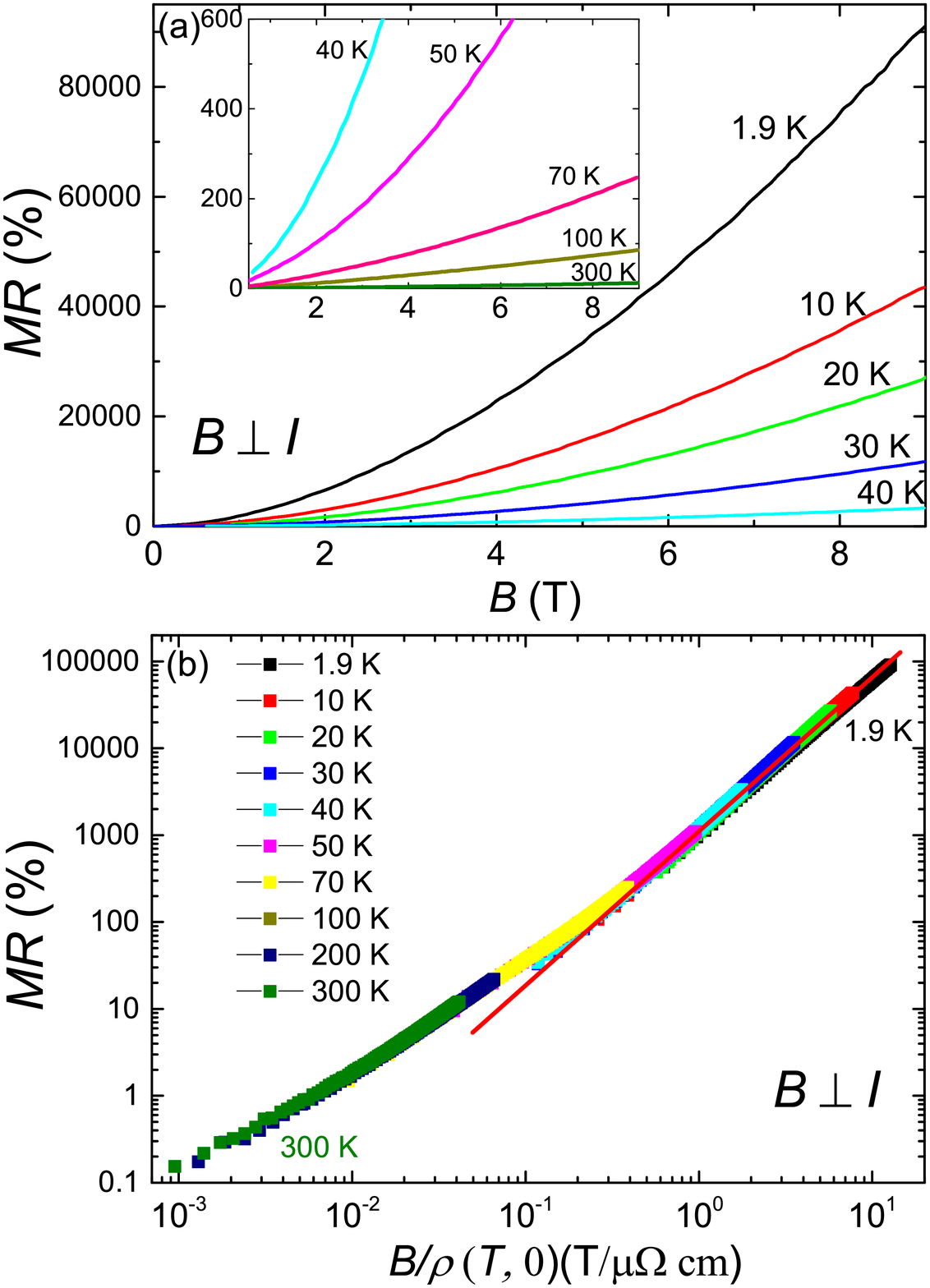}
\end{center}
	\caption{(Color online) (a) Transverse magnetoresistance of \\YbAs, measured at different temperatures from 300~K to 1.9~K in fields up to 9~T. (b) Kohler's rule scaling of the magnetoresistance as a function of \emph{B}/$\rho$(\emph{T}, 0).}
   \label{figure4}
\end{figure}

\begin{figure}[t]
\begin{center}
  \includegraphics[width=\columnwidth]{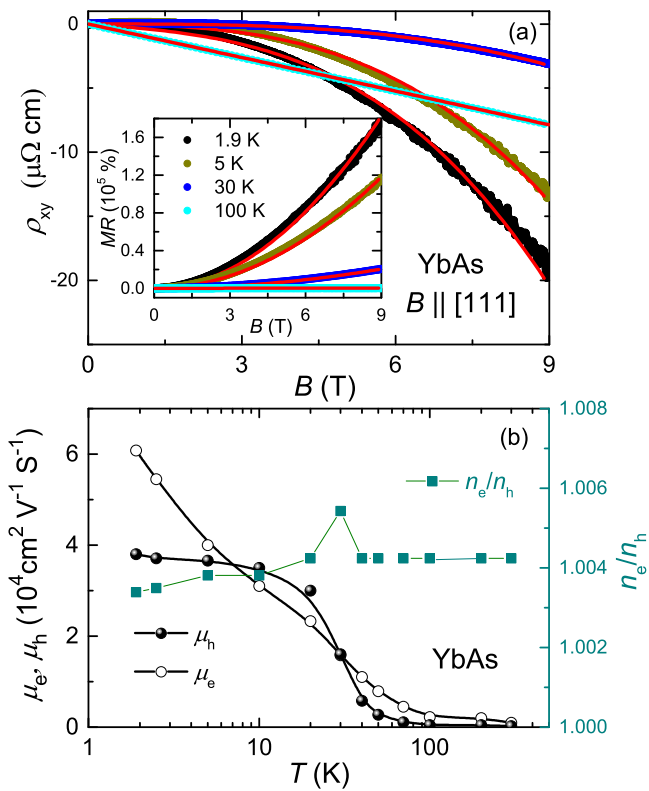}
\end{center}
	\caption{(Color online) (a) Field dependence of the Hall resistivity $\rho$$_{xy}$(\emph{B}) of YbAs, up to 9 T at 1.9, 5, 30, and 100~K. The inset shows the field dependent MR at the corresponding temperatures. The red solid lines in both the main panel and the inset show the results from fitting with a two-band model described in the text. (b) The temperature dependence of $n_e$/$n_h$, as well as the mobility of both electron- and hole- carriers, obtained from fitting both the MR and Hall resistivity with a two-band model. }
   \label{figure5}
\end{figure}

\subsection{MR and Hall resistivity}

In the main panel of Fig.~\ref{figure3}, $\rho$$_{xx}$(\emph{T}) under different applied magnetic fields is displayed for YbAs and LuAs down to 1.9~K. The field direction is perpendicular to the crystallographic plane and current direction, with $I \parallel [100]$ and $B \perp [100]$. As shown in the insets of Fig.~\ref{figure3}, $\rho_{xx}(T, 0)$ flattens below 10~K for LuAs, while $\rho_{xx}(T, 0)$ of YbAs has a change of slope at about 20~K, and decreases at a greater rate below 8~K, close to the temperature below which $C_\mathrm{mag}/T$ (Fig.~\ref{figure1}(c)) begins to increase more rapidly. The temperature dependent resistivity $\rho$$_{xx}$(\emph{T}, \emph{B}) of both YbAs and LuAs in field starts to upturn below a temperature denoted as $T\rm_{min}$, which increases with the applied field. For LuAs, a resistivity plateau develops below about 10~K, which is similar to other non-magnetic XMR materials such as La\emph{X} \cite{11-2015LaSb,11-2016NJP-LaBi,11-2017LaAs}. However, the in-field resistivity of YbAs keeps rising down to 1.9~K, and there is a clear change of slope at 20 K, together with a more rapid increase below around 8 K. As shown in Fig.~\ref{figure2}(a), the in-field $\rho$$_{xx}$(\emph{T}) continues to increase slightly at low temperatures, upon crossing the transitions at $T\rm_N$ or $T\rm_M$. Here we note that INS measurements show that anomalous short-range antiferromagnetic spin correlations appear below 20~K, and the INS peak intensity increases more strongly with decreasing temperature below about 8~K \cite{1995short range,1998short range}. These two temperatures correspond well to the anomalies observed in both the zero-field and in-field resistivity. The more rapid decreases in $\rho_{xx}(T, 0)$ upon lowering the temperature below 20 and 8 K (compared to that of LuAs) may be due to the reduction of spin-disorder scattering resulting from the development of short-range spin-correlations, which are enhanced below 8 K. This behavior in the low temperature $\rho_{xx}(T, 0)$ can give rise to the features observed in the in-field $\rho_{xx}(T, B)$ of YbAs across the same temperature range, as shown below.

To gain more insights into the observed XMR, the MR (= $100\%\times[\rho_{xx}(T, B)-\rho_{xx}(T, 0)]/\rho_{xx}(T, 0))$ of YbAs was analyzed to look for the presence of Kohler's rule scaling, as performed for other nonmagnetic XMR materials \cite{11-2016NJP-LaBi,11-2017LaAs,2015wang WTe2}. Figure~\ref{figure4}(a) displays the isothermal MR measured at various temperatures between 300~K and 1.9~K, which is plotted as a function of $\emph{B}$/$\rho$($\emph{T}$, 0) in Fig.~\ref{figure4}(b). It is apparent that the data at different temperatures fall onto a single line, showing that Kohler's rule is valid in YbAs \cite{kohler scaling}. Similar to WTe$_2$, this excludes the scenario of a metal-insulator transition being the origin of the field-induced resistivity upturn \cite{2015wang WTe2}. For WTe$_2$, the overall resistivity upturn behavior can be explained by Kohler's rule scaling,
 \begin{equation}
MR=\alpha(\emph{B}/\rho(\emph{T}, 0))^m,
\label{equation1}
\end{equation}
with \emph{m} $\simeq$ 2, which can be derived from a two-band model with electron-hole compensation \cite{2015wang WTe2}. For YbAs as shown in Fig.~\ref{figure4}(b), when the data are plotted on a double logarithmic scale, there is an apparent change in the slope at around 50--70 K, while below 50 K the data scale well linearly, with a fitted value of \emph{m} = 1.75.
Changes of the carrier concentration and/or mobility in the temperature range of 50--70 K may account for this change of slope \cite{2015wang WTe2,11-2016NJP-LaBi}.
The presence of Kohler's scaling shows that $\rho(T, B)$ is solely determined by $\rho$(\emph{T}, 0). This suggests that in YbAs the unsaturated temperature dependence of the in-field resistivity is a reflection of the continuous decrease of $\rho$(\emph{T}, 0) [inset of Fig.~\ref{figure3}(a)], which is most likely due to the reduced spin-disorder scattering arising from the development of
antiferromagnetic correlations in the corresponding temperature range.

The Hall resistivity $\rho$$_{xy}$(\emph{T}, \emph{B}) was also measured, so as to probe the temperature dependence of carrier concentration or mobility. As displayed in Fig.~\ref{figure5}(a), $\rho$$_{xy}$(\emph{B}) for \emph{T} $>$ 50 K exhibits a nearly linear field dependence, while it is non-linear below 50 K.

\begin{figure*}[t]
\begin{center}
  \includegraphics[width=\textwidth]{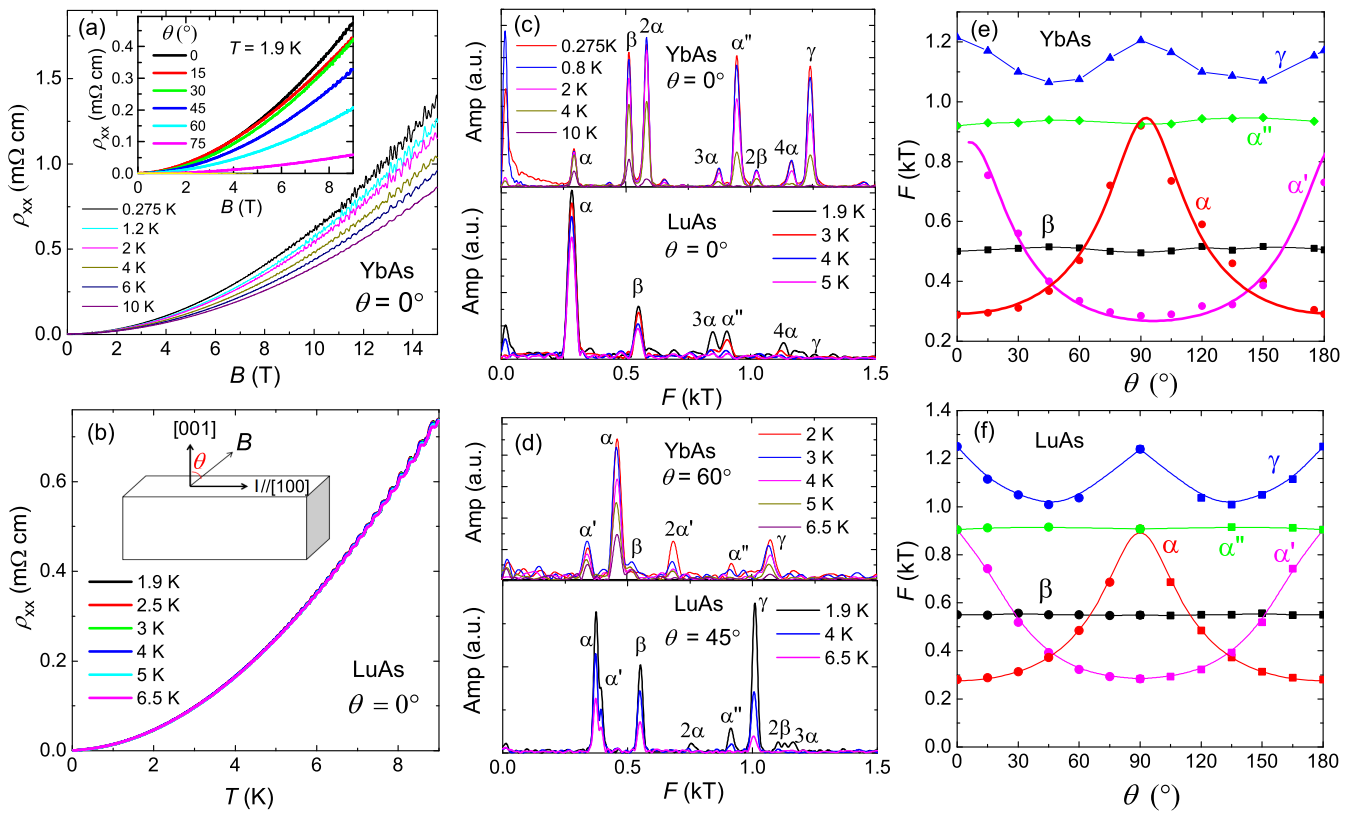}
\end{center}
	\caption{(Color online) Magnetic field dependence of the resistivity($\rho$$_{xx}$) of (a) YbAs, and (b) LuAs. The inset of (a) shows the field dependence of $\rho_{xx}$ of YbAs at different $\theta$ values at 1.9~K, where $\theta$ is the angle between the field direction and [001], as illustrated in the inset of panel (b). (c) Fast-Fourier-transform (FFT) analysis of the SdH oscillations at $\theta$ = 0$^\circ$ for both YbAs (field range of 6--15 T) and LuAs (field range of 6--9 T), respectively. (d) FFT results at $\theta$ = 60$^\circ$ for YbAs and $\theta$ = 45$^\circ$ for LuAs, respectively, both in a field range of 6--9 T. (e) SdH oscillation frequencies as a function of $\theta$ for YbAs, obtained via an FFT analysis on the angle-dependent MR. The fitted curves described in the main text are shown by the thick lines for the $\alpha$ and $\alpha$$'$ bands, while the other thin lines are guides to the eyes. (f) SdH oscillation frequencies as a function of $\theta$ for LuAs. The thin solid lines are guides to the eyes. The frequencies \emph{F} for $\theta$ $>$ 90$^\circ$ are mirrored from the 0--90$^\circ$ data, in order to compare with YbAs.}
   \label{figure6}
\end{figure*}

For a two-band model \cite{kohler scaling,11-2016NJP-LaBi}, \emph{$\rho$}$_{xy}$(\emph{B}) and the MR can be expressed as
\begin{equation}
 \rho_{xy}=\frac{B}{e}
 \frac{(n_h\mu_h^2-n_e\mu_e^2)+(n_h-n_e)(\mu_e\mu_h)^2B^2}{(n_h\mu_{h}+n_e\mu_{e})^2+(n_h-n_e)^2(\mu_{e}\mu_{h})^2B^2},
\label{}
\end{equation}
\begin{equation}
 MR=\frac{n_e\mu_en_h\mu_h(\mu_e+\mu_h)^2B^2}{(n_h\mu_{h}+n_e\mu_{e})^2+(n_h-n_e)^2(\mu_{e}\mu_{h})^2B^2}.
\label{}
\end{equation}
where $n_e$ and $n_h$ are the carrier concentrations of electrons and holes, respectively, while $\mu_e$ and $\mu_h$ are the respective mobilities.

Both $\rho$$_{xy}$(\emph{B}) and the MR were simultaneously fitted using Eqs. (2) and (3), and the fitted curves at several temperatures are displayed in Fig.~\ref{figure5}(a). Due to the large number of parameters, $n_e$ was fixed to the value of 4.74$\times$10$^{20}$~cm$^{-3}$ from the analysis of the SdH oscillations (see below). Here we note that while this model is relatively insensitive to the absolute values of $n_e$ and $n_h$, the ratio $n_e$/$n_h$ is  well constrained.  In Fig.~\ref{figure5}(b), the temperature dependence of the fitted values of $n_e$/$n_h$, $\mu_e$, and $\mu_h$ are displayed. It can be seen that both $\mu_e$ and $\mu_h$ show a significant increase upon cooling below about 70 K, which may explain the change of exponent in the Kohler's rule scaling. The large values of the mobilities, reaching 4--6 m$^2$/(V$\cdot$s) at 1.9 K, can also account for the large magnetoresistance at low temperatures. Note that these mobilities are quite different from previous results \cite{1993mobility}, where only the electron mobility displays a weak increase below about 80 K, and the values are an order of magnitude lower. We ascribe this difference to the higher quality of our single crystals, manifested in the much larger RRR value. The fitted electron- and hole- carrier densities are nearly compensated over the whole temperature range ($n_e$/$n_h$ $\approx$ 1), in agreement with previous reports \cite{1993mobility}. Therefore the combination of the validity of Kohler's rule scaling and the  analysis of the magnetotransport with a two-band model indicates that electron-hole compensation is the main mechanism for the observed XMR \cite{11-2016NJP-LaBi,11-2017LaAs,2015wang WTe2}. In particular, the large mobilities at low temperatures for both electrons and holes could help realize the XMR in YbAs \cite{11-2017LaAs}, but is different to that expected for the $p-f$ mixing model, which predicts that the mobility of the electrons is much larger than the holes \cite{1990transport,1993mobility}.

\subsection{Fermi surfaces}

\subsubsection{SdH oscillations}

\begin{figure}[t]
\begin{center}
  \includegraphics[width=0.8\columnwidth,keepaspectratio]{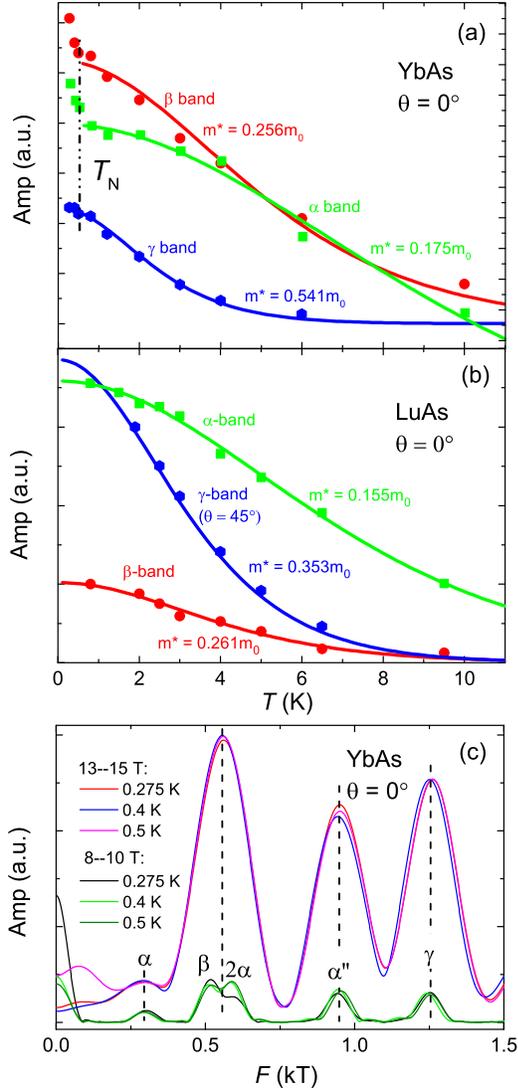}
\end{center}
	\caption{(Color online)  Temperature dependence of the SdH oscillation amplitudes of all three bands for (a) YbAs (field range of 6--11 T) and (b) LuAs (field range of 6--9 T). The solid lines show the results of fitting with the L-K formula. The effective masses for the bands are labelled. (c) FFT analysis for YbAs at three temperatures, both above and below $T\rm_N$ or $T\rm_M$, which show negligible differences in the oscillation frequencies.}
   \label{figure7}
\end{figure}

To gain more insight into the electronic structure and to examine the band topology of YbAs and LuAs, we studied the Fermi surfaces using angle-dependent SdH oscillations measurements. Clear SdH oscillations are observed in the MR of YbAs at temperatures up to $\sim$ 10 K in fields above 6~T, and to further examine the low temperature behavior, MR measurements in fields up to 15~T and down to 0.275~K were performed, and the results are displayed in the main panel of Fig.~\ref{figure6}(a).
 Meanwhile the data for LuAs down to 1.9~K and up to 9~T are shown in Fig.~\ref{figure6}(b).  Angle dependent $\rho$$_{xx}$ measurements were also performed for both YbAs [inset of Fig.~\ref{figure6}(a)] and LuAs, for different angles $\theta$ between the applied field and the [001] direction, as illustrated in the inset of Fig.~\ref{figure6}(b). After subtracting the background,  the SdH oscillations were analyzed via  fast-Fourier-transforms (FFT)  for various values of $\theta$, and the results are shown in Fig.~\ref{figure6}(c) and Fig.~\ref{figure6}(d).
 For both compounds, we find the presence of five principal frequencies ($\alpha$, $\alpha$$'$, $\alpha$$''$, $\beta$, and $\gamma$ ) and some harmonic frequencies, indicating multiple Fermi surface pockets. From comparing the FFT results between YbAs and LuAs at $\theta$ = 0$^\circ$, it can be seen that the frequencies of the $\alpha$ and $\gamma$ pockets ($F_{\alpha}$ and $F_{\gamma}$) are very close, while the frequency of $\beta$ pockets ($F_{\beta}$) of LuAs is slightly larger than that of YbAs and $F_{\alpha''}$ is slightly smaller. Note that $\alpha$$''$ coincides with $\alpha$$'$ at $\theta$ = 0$^\circ$, while they are separated at $\theta$ = 45$^\circ$ and $\theta$ = 60$^\circ$ [Fig.~\ref{figure6}(d)]. The FFT spectra of YbAs at 0.275 K are identical to those at 10 K, indicating that the Fermi surface does not change with temperature in this range.

The respective angle dependences of the SdH frequencies are displayed in Fig.~\ref{figure6}(e) and Fig.~\ref{figure6}(f), obtained from performing an FFT analysis on data at different angles in the field range of 6--9 T. The lines correspond to the five pockets, where the cross-sectional areas of the extremal orbits ($A\rm_{ext}$) change upon rotating the field direction according to the Onsager relation \emph{F} $\equiv$ $(\Phi_0$/2$\pi^2$)$\times$$A\rm_{ext}$, where $\Phi$$_0$ is the magnetic flux quantum. The dispersions shown here are consistent with the dHvA results reported for YbAs in Ref. \cite{1992dHvA} and similar to those of LaAs \cite{11-2017LaAs}. In Ref. \cite{1992dHvA}, four angle-dependent dispersion lines were observed for YbAs, three ellipsoidal electron pockets and only one spherical hole pocket, which correspond to the $\alpha$, $\alpha$$'$, $\alpha$$''$ pockets and the $\beta$ pocket in our data, respectively. The presence of at least one more hole pocket with a large effective mass was proposed based on the \emph{p}-\emph{f} mixing model \cite{1992dHvA}. Here we find that an extra hole pocket is indeed present (labelled as $\gamma$) for both YbAs and LuAs, which likely corresponds to a jack shape, as observed in LaAs \cite{1992dHvA,11-2017LaAs}. The similar angle-dependent dispersions of YbAs to that of LuAs indicates that their Fermi surface geometries are nearly identical, implying that the Yb-4\emph{f} electrons are well localized across the measured temperature range.

We calculated the carrier concentrations $\emph{n}$ of both electrons and holes, which are proportional to the volume $\emph{V}$$\rm_F$ of the Fermi pockets via \emph{n} = \emph{V}$\rm_F$/4$\pi$$^3$. Here the angular dependence of $A_{ext}$ was used to extract the lengths of the semi-major and semi-minor axes of the ellipsoidal pockets, as well as the radius of the spherical pocket. For the irregularly shaped $\gamma$-pocket, $V_F$ was estimated following the method used for LaAs in Ref. \cite{11-2017LaAs}.
The calculated carrier densities are \emph{n}$_e$ = 4.74$\times$10$^{20}$~cm$^{-3}$ and \emph{n}$_e$/\emph{n}$_h$ = 1.082 for YbAs, and \emph{n}$_e$ = 4.56$\times$10$^{20}$~cm$^{-3}$ with \emph{n}$_e$/\emph{n}$_h$ = 0.98 for LuAs. The electron-hole compensation is consistent with the analysis of the MR and Hall resistivity, and moreover indicates that all the Fermi surface pockets in the Brillouin zone were detected from our SdH measurements.

For three-dimensional ellipsoidal pockets, the angular dependent quantum oscillation frequency \emph{F}$(\theta)$ can be described by \cite{2012-ellipsoid}

\begin{equation}
F(\theta)=A\pi ab/\sqrt{\sin^2({\theta+\varphi})+(a^2/b^2)\cos^2({\theta+\varphi})},
\label{equation2}
\end{equation}

 \noindent where \emph{a} and \emph{b} are the lengths of the semi-major axis and semi-minor axis of the ellipsoidal $\alpha$ pockets, respectively, \emph{A} is a constant, and $\varphi$ is an angular shift \cite{2012-ellipsoid}. By using the above derived parameters \emph{a} and \emph{b}, it is shown in Fig.~\ref{figure6}(e) that the angular dependence of the $\alpha$ and $\alpha$$'$ band frequencies of YbAs can be well accounted for by Eq. (4), with fitted values of \emph{A} = $ 3.35\times10^{4}  \rm{\AA}^2 T$ and $\varphi$ = --1.26$^\circ$.

The temperature dependence of the amplitudes for all the three bands at $\theta = 0^\circ$ were extracted, which were fitted using the Lifshitz-Kosevich (L-K) formula \cite{LK formula}, as shown in Fig.~\ref{figure7}(a). For the $\gamma$ band, this yields an effective mass of 0.541$\emph{m}$$\rm_0$ ($\emph{m}$$\rm_0$ is the free electron mass) and no anomaly is observed upon crossing \emph{T}$\rm_N$.  However, for the $\alpha$ and $\beta$ bands, which have lighter effective masses of 0.175$\emph{m}$$_0$ and 0.256$\emph{m}$$_0$, respectively, an abrupt increase of the amplitude is observed upon cooling below \emph{T}$\rm_N$. The effective mass of the $\alpha$-band is consistent with previously reported dHvA-effect mesurements \cite{1992dHvA}, where the increase in amplitude of the $\alpha$ band below \emph{T}$\rm_N$  was explained as arising due to the combined effects of the spin factor and a change of Dingle temperature \cite{1992dHvA}. For LuAs, as shown in Fig.~\ref{figure7}(b), the obtained effective masses for the $\alpha$, $\beta$ and $\gamma$ bands are 0.155, 0.261 and 0.353$\emph{m}$$\rm_0$ (for $\theta = 45^\circ$), respectively. The results are summarized in Table. I for comparison, where the effective masses obtained from DFT-calculations for YbAs are also listed, assuming localized \emph{f}-electrons. This shows that similarly small effective charge carrier masses are deduced for YbAs and LuAs, which are well accounted for by the DFT calculations. The above analysis indicates that YbAs has a similar Fermi surface to LuAs in the studied temperature range from 10 K down to 0.275 K. The carrier effective  masses of YbAs are not significantly enhanced compared to those of LuAs, indicating that down to at least \emph{T}$\rm_N$ the 4\emph{f}-electrons are well localized in YbAs, and the electronic correlations do not lead to a large renormalization of carrier masses.

To examine the possible influence of the transitions at \emph{T}$\rm_N$ and \emph{T}$\rm_M$ on the Fermi surface, we analyzed the FFT in several temperature and field ranges, corresponding to different phases in the $\emph{B}$--$\emph{T}$ phase diagram shown in Fig.~\ref{figure2}(b). The results displayed in Fig.~\ref{figure7}(c) indicate that the SdH frequencies are nearly unchanged, suggesting that the Fermi surface volume is not modified upon crossing \emph{T}$\rm_N$ or \emph{T}$\rm_M$.\\

\begin{table}
  \centering
  \caption{Comparison of effective carrier masses for the three bands obtained experimentally for YbAs and LuAs, together with the results from DFT calculation of YbAs.  Besides those labelled with brackets, the values are taken for $\theta = 0^\circ$. }
  \begin{tabular}{p{0.5in}p{0.5in}p{0.5in}c}
  \hline\hline
    & $\alpha$ & $\beta$ & $\gamma$ \\
    & (m$_0$) & (m$_0$) & (m$_0$)\\ [0.5ex]
   \hline

    YbAs  & 0.175  & 0.256 & 0.541 \\ %[0.5ex]
    \hline
    LuAs & 0.155 & 0.261 & 0.353($45^\circ$)  \\ %[-1ex]
    \hline

    DFT & 0.147 & 0.204 & 0.427, 0.350(45$^\circ$)  \\ %[-1ex]

   \hline\hline
   \end{tabular}
\end{table}

\begin{figure}[t]
\begin{center}
  \includegraphics[width=\columnwidth]{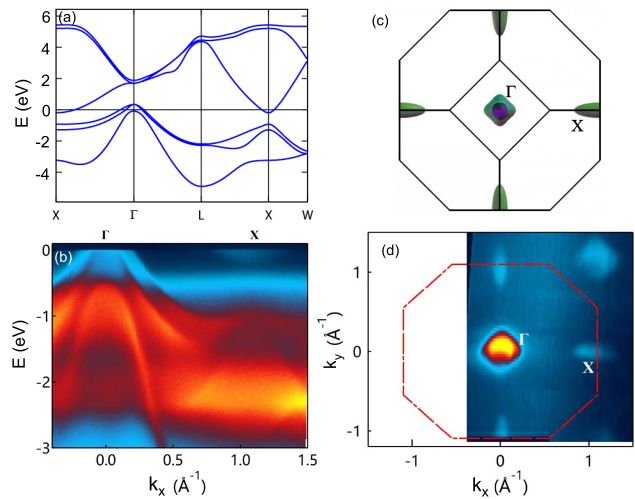}
\end{center}
	\caption{(Color online) (a) The calculated band structure of YbAs along high-symmetry lines. (b) ARPES energy-momentum cut along the \emph{k$_x$} direction ($\Gamma$-X) at \emph{k$_y$}~=~0 and \emph{k$_z$} $\approx$ 0, to be compared with the $\Gamma$-X direction in (a). (c) The calculated 3D Fermi surfaces at \emph{k$_z$}~=~0 viewed from the \emph{z}-direction. (d) Experimental Fermi surface in the \emph{k}$_x$-\emph{k}$_y$ plane near \emph{k$_z$}~=~0. The photon energy used for (b) and (d) is 60 eV, which corresponds to the \emph{k$_z$} = 0 cut. The broken red line sketches the Brillouin Zone (BZ). }
   \label{figure8}
\end{figure}

\begin{figure}[t]
\begin{center}
  \includegraphics[width=\columnwidth]{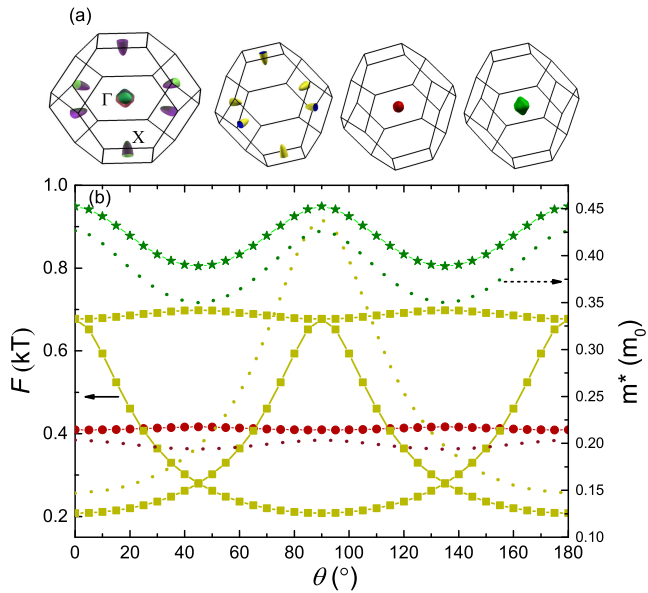}
\end{center}
	\caption{(Color online) (a) The calculated Fermi surfaces of YbAs, including electron pockets (dark yellow) at the X points and hole pockets (dark red and dark green) around the $\Gamma$ point in the Brillouin zone. (b) The calculated angle dependent dispersions of the FFT frequencies based on the Fermi pockets shown in (a) and the corresponding effective masses of the respective bands (shown by dotted lines).  }
   \label{figure9}
\end{figure}

\begin{figure}[t]
\begin{center}
  \includegraphics[width=\columnwidth]{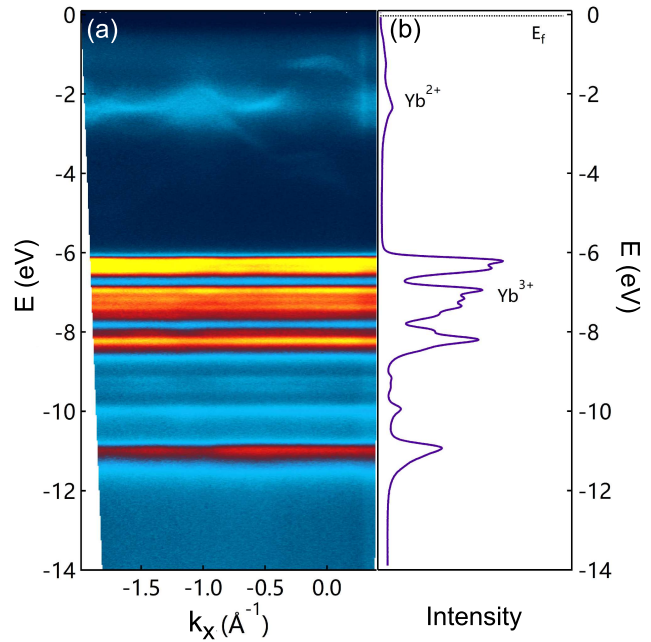}
\end{center}
	\caption{(Color online) (a) Large energy scale ARPES spectrum along the \emph{k$_x$} direction at \emph{k$_y$} = 0 taken with 90 eV photons and (b) the angle-integrated energy distribution curves showing the spin-orbit split Yb$^{4+}$ 4\emph{f} states at -1 and -2 eV, and multiple Yb$^{3+}$ 4\emph{f} states between -6 and -12 eV.}
   \label{figure10}
\end{figure}

\subsubsection{ DFT calculations and ARPES measurements}

To further study the Fermi surfaces and topology of YbAs, we performed DFT calculations and ARPES measurements down to 10 K. In Figs.~\ref{figure8}(a) and (c), we show the calculated band structure and Fermi surface of YbAs. Figure~\ref{figure8}(b) displays the ARPES energy vs momentum cut along the \emph{k$_x$} direction at \emph{k$_y$} = 0 and \emph{k$_z$} $\approx$ 0, while Fig.~\ref{figure8}(d) shows the Fermi surface map in the \emph{k$_x$}-\emph{k$_y$} plane at \emph{k$_z$} $\approx$ 0. The ARPES spectra were taken at 60 eV, which corresponds to the \emph{k$_z$} = 0 condition.  From comparing the calculated results with the measured data, it is clear that the Fermi surfaces consist of two concentric hole pockets at the $\rm$$\Gamma$ point and three symmetry-equivalent electron pockets at the X points, which are consistent with the results of our FFT analysis. In general, the ARPES spectra are in good agreement with the DFT calculations, indicating that the \emph{f} electrons are mostly localized. In Figs.~\ref{figure8}(a) and (b), no band inversion along the $\rm$$\Gamma$-X direction can be observed, unlike the rare-earth monobismuthides \cite{11-2018PLi}, indicating the topologically trivial nature of YbAs, similar to LaAs \cite{11-2017LaAs}. This could be a consequence of  the weaker spin-orbit coupling induced by As compared to Bi.

To highlight the different parts of the Fermi surface, we show the separated pockets in Fig.~\ref{figure9} together with the calculated angle-dependent dispersions of the FFT frequencies. These dispersions agree well with our FFT analysis displayed in Fig.~\ref{figure6}(d), while the systematic differences in the values of the frequencies may arise from the presence of possible correlations in YbAs. Furthermore, the angle dependent effective masses of the respective bands are also calculated, shown in Fig.~\ref{figure9}(b) as dotted lines. This further indicates that the effective masses are small along all directions for all bands, with the values at $\theta = 0^\circ$ reasonably consistent with the experimental results.

Furthermore, Fig.~\ref{figure8}(b) shows that no coherent bulk 4\emph{f} band, i.e., the Kondo resonance peak, can be observed near the Fermi level, which indicates that the heavy fermion state has not developed down to 10~K. To confirm that the absence of a Kondo resonance peak is not due to a diminishing photoexcitation cross section, we employed a large range of photon energies (from 30 eV up to 200 eV) to search for any weak signature of a Kondo resonance peak near the Fermi level. Nevertheless, all our ARPES spectra show similar flat \emph{f} bands at --1 and --2.1 eV (as shown in Fig.~\ref{figure8}(b)), which are far away from the Fermi level and therefore are different from the bulk Kondo resonance that always resides near the Fermi level. A large-energy-range spectrum with its angle-integrated energy distribution curve is shown in Fig.~\ref{figure10}, which reveals multiple intense Yb$^{3+}$ peaks ranging from --12 eV to --6 eV, and weak Yb$^{2+}$ peaks at --1 and --2.1 eV, respectively. The sharp Yb$^{3+}$ peaks confirm that the valence of Yb is mainly trivalent in YbAs, while the weak peaks at --1 and --2.1 eV are most likely due to surface Yb$^{2+}$ contributions, as seen in many Yb-based heavy fermion compounds \cite{2016arpes}. Temperature dependent scans were also performed from 13 K up to 80 K, but no signature of a Kondo resonance peak near the Fermi level can be observed and the band structure shows essentially no temperature dependence. This indicates that the many-body Kondo screening is rather weak at temperatures of 10 K and above.

\section{SUMMARY}

In summary, we have grown high-quality single crystals of YbAs and performed a detailed study of the magnetotransport and ARPES measurements,  together with DFT calculations. From low temperature resistivity measurements, we constructed a temperature--field phase diagram of YbAs for $B \perp $ [100], where we find a different field-induced first-order phase transition at high fields ($\geq$ 11~T). Moreover, XMR is observed in YbAs, and our analysis of the MR and Hall resistivity indicates that electron-hole compensation with enhanced carrier mobilities is the main mechanism leading to this behavior. However, upon comparing with the MR data of LuAs, distinct anomalies are discerned at low temperatures in YbAs, which are attributed to the short-range magnetic correlations revealed by INS measurements \cite{1995short range,1998short range}.

SdH oscillations for YbAs and LuAs reveal the presence of three electron and two hole pockets, where light effective carrier masses are deduced from an analysis of the temperature dependence of the oscillation amplitudes down to 0.5 K.  Together with ARPES measurements and DFT calculations, we found that the Fermi surface of YbAs is similar to that of non-magnetic  LuAs, indicating that the 4\emph{f}-electrons are well localized over the analyzed temperature range. Since the enhanced Sommerfeld coefficient and nuclear relaxation rate $1/(T_1T)$ were deduced from measurements below 0.3~K \cite{Takuo1992,1998NMR}, it is possible that the system only displays heavy fermion behavior at temperatures lower than the present study. The magnetic ordering temperature of YbAs ($T\rm_{N}$ = 0.5 K) is much lower than the estimated energy scale of the magnetic exchange interactions ($\approx$ 10 K) from M\"{o}ssbauer spectroscopy \cite{Bonville1988}, which was originally attributed to the presence of a significant Kondo effect \cite{Bonville1988,1988Bonville et al,1988Bonville}. However, we note that type-III AFM order on a face-centered cubic lattice, as in the case of YbAs, can also exhibit strong magnetic frustration \cite{MnS2}, which could be the origin of both the low $T\rm_{N}$ and the existence of short-range correlations at elevated temperatures.

Finally,  ARPES and DFT calculations reveal a lack of band inversions in YbAs, indicating a topologically trivial electronic structure. While therefore YbAs does not exhibit the coexistence of magnetism and topological band structures, our results show that the XMR is significantly influenced by the presence short ranged magnetic correlations arising from localized electronic moments.

\section{Acknowledgments}
We thank  C. Y. Guo, T. Shang, L. Jiao for their helpful suggestions. This work was supported
by the National Key R$\&$D Program of China (Grants No. 2017YFA0303100 and No. 2016YFA0300202), the National
Natural Science Foundation of China (Grants No. U1632275, No. 11604291 and No. 11974306), and the Science Challenge Project of
China (Project No. TZ2016004).
This research used resources of the Advanced Light Source, which is a DOE Office of Science User Facility under contract No. DE-AC02-05CH11231. Y. Wu thanks Dr. Peng Chen for the assistance in the ARPES experiments.

\end{document}